\documentstyle[12pt,a4,thmsa,sw20jar2]{article}

\input{tcilatex}
\input tcilatex
\begin{document}

\title{Speed of light on rotating platforms}
\author{G. Rizzi, A. Tartaglia \\
Dip. Fisica, Politecnico, Corso Duca degli Abruzzi 24, I-10129 Turin, Italy\\
E-mail tartaglia@polito.it}
\maketitle

\begin{abstract}
It is often taken for granted that on board a rotating disk it is possible
to operate a {\it global }3+1 splitting of space-time, such that both
lengths and time intervals are {\it uniquely} defined in terms of
measurements performed by real rods and real clocks at rest on the platform.
This paper shows that this assumption, although widespread and apparently
trivial, leads to an anisotropy of the velocity of two light beams
travelling in opposite directions along the rim of the disk; which in turn
implies some recently pointed out paradoxical consequences undermining the
self-consistency of the Special Theory of Relativity (SRT). A correct
application of the SRT solves the problem and recovers complete internal
consistency for the theory. As an immediate consequence, it is shown that
the Sagnac effect only depends on the non homogeneity of time on the
platform and has nothing to do with any anisotropy of the speed of light
along the rim of the disk, contrary to an incorrect but widely supported
idea.
\end{abstract}

\section{Introduction}

After almost one century of Relativity, many pre-relativistic prejudices
(Baconian ''idola'') still survive. One of the most tenacious is the idea
that, after giving up Newton's absolute space and absolute time (the 3+1
absolute splitting), every observer (or ''reference frame'') possesses in
any case its own private space and its own private time, i.e. its private 
{\it extended }3+1 splitting. However, it should be known from the General
Theory of Relativity that this can be true only for the class of the {\it %
extended reference frames defined by a congruence }$\Gamma ${\it \ of
timelike worldlines, living in general Riemannian space-times, for which the
''vortex tensor'' \cite{cattaneo} vanishes}\footnote{%
In this case the reference frame is said to be {\it time-orthogonal }and 
{\it geodesic} \cite{cattaneo}. An obvious (but not trivial) example is a
non-accelerated physical frame in a static gravitational field, for instance
in a Schwarzschild space-time.}. Let us point out that this class includes
the important subclass of the {\it extended inertial frames} {\it living in
Minkowskian space-times} (space-times with vanishing Riemann tensor).

As a consequence, in any reference frame for which the vortex tensor differs
from zero, the concept of ''the whole physical space at a given instant''
turns out to be conventional, in the sense that it is lacking an operational
meaning because of the impossibility of a symmetrical {\it and transitive}
synchronization procedure at large\footnote{%
Such procedure is possible only locally, according to the principle of local
equivalence between an accelerated observer and an instantaneously comoving
inertial observer. In other words, the topology of spacetime insures the
possibility of a {\it local (but only local) }3+1 splitting.}.

In this paper, we shall deal with a very interesting case, where the naive
assumption of the existence of a ''physical space'' in a reference frame for
which the vortex tensor is different from zero leads to paradoxical results.
This is the case, widely studied in the literature but often plagued by
serious misunderstandings, of a disk uniformly rotating in a Minkowskian
space-time. This problem has been treated by various authors with different
approaches (the difference being essentially in the definition of space and
time on the disk), all of them (with only a few exceptions, like Cantoni 
\cite{cantoni}, Anandan\cite{anandan} and Mashhoon\cite{mashhoon}) sharing a
crucial point which, as we shall see, contains a fundamental element of
ambiguity: {\it the circumference of the disk is treated as a geometrically
well defined entity, that possesses a well defined length}\footnote{%
An example is the approach by Landau and Lifshitz who move from the
(apparently trivial) remark: ''Let us consider two reference frames one of
them (K) being inertial, the other (K') uniformly rotating with respect to K
about the common z axis. A circumference on the xy plane of reference K
(centered at the origin of the coordinates) can be considered as a
circumference in the x'y' plane of reference K'.'' (cfr. \cite{landau}, \S\
82).} without worrying about the fact that no transitive synchronism exists
along the said circumference. Further on they diverge: (i) on the measure of
such a length; (ii) on the time unit used to evaluate the velocity of
(massive or massless) particles in uniform motion along the said
circumference (we do not consider here other essentially derived topics,
such as for instance the space metric on the disk).

To fix ideas, let $K$ be an inertial frame, and $K_{o\text{ }}$a rigid
circular platform rotating with constant angular velocity $\omega $ with
respect to $K$ (in the following, all quantities valued in $K_o$ will be
indicated by the suffix $_o$; all quantities valued in $K$ will be indicated
without any suffix). Almost all authors consider two circumferences: (i) the
rim of the platform, as seen in $K_o$; (ii) the set of the positions of the
points of the rim in $K$; and assume that these circumferences are
geometrically (of course not kinematically) identical. Let $R_o$, $R$, be
the lengths of the radii of the two circumferences, as seen in $K_o$, $K$
respectively; $L_o$, $L$ the lengths of the circumferences; $t_o$ an
interval of proper time of a clock $C_\Sigma $ at rest on the rim, and $t$
the corresponding interval of time in $K$, as measured by two different
clocks at rest in $K$, and there synchronized.

The relationships between the above quantities have the very general form

\begin{eqnarray}
R &=&R_oF_R(v,a)  \label{rlt} \\
L &=&L_oF_L(v,a)  \nonumber \\
t &=&t_oF_t(v,a)  \nonumber
\end{eqnarray}
where $F_R$, $F_L$, $F_t$ are generic functions of the peripheral velocity $%
v=\omega R$ (which of course is assumed to be lower than $c$) and possibly
of the centrifugal acceleration $a=v^2/R=\omega ^2R$ (although the
dependence from the acceleration is not expected, in the standard theory of
Relativity).

According to the Special Theory of Relativity (SRT), all authors agree about 
$F_R$, $F_L$ and $F_t$:

\begin{equation}
F_R=1  \label{fl0}
\end{equation}

\begin{equation}
F_L=\gamma ^{-1}(v)\equiv \left( 1-\beta ^2\right) ^{1/2}  \label{fl1}
\end{equation}

\begin{equation}
F_t=\gamma (v)\equiv \left( 1-\beta ^2\right) ^{-1/2}  \label{fl2}
\end{equation}
($\beta \equiv v/c$).

The assumptions (\ref{fl0}), (\ref{fl1}), (\ref{fl2}) come from the
consideration that: (i) the Lorentz contraction acts only on the periphery
and not on the radius of the disk; (ii) any clock $C_\Sigma $ at rest on the
rim of the disk undergoes the Lorentz dilation of time intervals, as
measured in $K$.

However, not all authors agree about the numerical value of the ratios $%
L_o/R_o$ and $L/R$. As a matter of fact, two different assumptions are found
in the literature:

\begin{equation}
\frac{L_o}{R_o}=2\pi  \label{ratio1}
\end{equation}
\begin{equation}
\frac LR=2\pi  \label{ratio2}
\end{equation}

The assumption (\ref{ratio1}) - see for instance \cite{cavalleri} and
references therein - comes from the consideration that, since the proper
length $dL_{o}$ of an infinitesimal element of the rotating circumference
does not change when the disk passes from rest to rotation (all proper
quantities are invariant), {\it the same should happen for the entire
circumference: this gives }$L_{o}=2\pi R_{o}=2\pi R${\it , which is
interpreted as the proper length of the circumference.} A puzzling, but
unavoidable, consequence is the following: the ratio $L/R$, as measured in $K
$, is less than $2\pi $, thus violating the Euclidean geometry of $K$
(remember that $K$ is an inertial frame!). This is the well known Ehrenfest
paradox \cite{ehrenfest}. The only way to maintain the Euclidean geometry of 
$K$, when $L<2\pi R${\it , }consists in introducing a further ad hoc
hypothesis, and precisely the hypothesis that the surface of the disk bends,
in a suitable way, because of the rotation. If such ad hoc hypothesis is
rejected, on the basis both of kinematical and dynamical considerations%
\footnote{%
The kinematical consideration is the following: the bending of the rotating
disk, obviously not-symmetric with respect to the plane of the disk when it
is not rotating, determines a skew sense in space, thus violating the
spatial parity of $K$ {\it on a purely kinematical basis}.}, the Ehrenfest
paradox cannot be solved, ''from a purely kinematic point of view'': this is
the conclusion of Cavalleri \cite{cavalleri}, who ends up with the statement
that ''the relativistic kinematics for extended bodies is not generally
self-consistent'' (and suggests an ''intrinsically dynamical'' solution of
the paradox such as the one invoked by Dieks\cite{dieks}).

The assumption (\ref{ratio2}) comes from the consideration that the edge $%
\gamma _{o\text{ }}$of the platform, as seen in $K_o$, and the set $\gamma $
of the positions of the points of such edge in $K$, {\it are two
circumferences geometrically }(although of course not kinematically) {\it %
identical}. Moreover: (i) the circumference $\gamma $, as seen in $K$, must
have the length $L=2\pi R$ , according to the Euclidean geometry of $K$;
(ii) since the circumference $\gamma _o$, as seen in $K_o$, is not changed,
but the unit (infinitesimal) rod is changed by a factor $\gamma ^{-1}(v)$
(Lorentz contraction), then the length of the circumference $\gamma _{o\text{
}}$, as measured in $K$, turns out to be increased by a factor of $\gamma $.
Therefore $L_o=2\pi R\gamma (v)>2\pi R$ : which shows that the geometry of
the rotating disk is not Euclidean. This is the most widespread assumption
(see for instance Einstein \cite{einstein}, Arzeli\`{e}s \cite{arzelies},
Landau and Lifshitz \cite{landau}, M\o ller \cite{moller}, etc.).

We are not going to comment further on these topics, but simply remark that
the widespread and apparently innocent assumption that the length of a round
trip along the border of the rotating disk coincides with that of a
univocally defined geometric object, unavoidably leads to the paradox
pointed out by Selleri \cite{selleri} whose case we shall summarize in the
next section. A great many authors did not realize this fact.

We shall show that a correct and thorough application of the Special Theory
of Relativity dissipates any ambiguity, by giving up the prejudice of the
unicity of the length of a round trip about the axis of a rotating turntable.

\section{Selleri's paradox}

Recently, an interesting paper by Selleri \cite{selleri} shows that:

a) {\it under the only }(apparently obvious and, as a matter of fact,
universally shared){\it \ assumption that a round trip corresponds to a
geometrically well defined entity, whose equally well defined length be }$%
L_o $ (about its measure any hypothesis is anyhow avoided){\it ;}

b) {\it independently of any of the particular assumptions mentioned in the
introduction; }

c) {\it more generally, independently of any particular choice about the
functions }$F_L$ {\it and }$F_t${\it ; }

{\it an unavoidable paradox actually arises in the standard Special
Relativity Theory applied to rotating platforms}.

Selleri considers a light source $\Sigma $ placed in a fixed position near
the clock $C_\Sigma $ at rest on the rim of the rotating disk, and two light
flashes leaving $\Sigma $ in opposite directions at the (proper) time $t_{o1%
\text{ }}$of $C_\Sigma $ , grazing a cylindrical mirror on the rim. The
counter-rotating and the co-rotating light signals come back to $\Sigma $
respectively at the (proper) times $t_{o2\text{ }}$and $t_{o3}$ of $C_\Sigma 
$ (these times are different because the light signals, as seen in the
inertial frame $K$, travel with the same velocity $c$ along two paths which
are different because of the rotation of the disk). On the other hand,
according to the previous assumption, the two paths should be identical,
with the same length $L_o$, in the frame $K_{o\text{ }}$of the disk. Then
the velocities $c_{o-}$, $\,c_{o+}$ of the counter-rotating and co-rotating
light signals, as seen in $K_o$, should be different:

\begin{equation}
c_{o-}=\frac{L_o}{t_{o2}-t_{o1}}\quad ;\quad \,\,\,\,c_{o+}=\frac{L_o}{%
t_{o3}-t_{o1}}  \label{c}
\end{equation}
(since $K_o$ is not an inertial frame, ''there is no reason to demand that
the speed of light be the same eastward and westward''\cite{peres}). If $%
c_{o-}$, $\,c_{o+}$ are expressed in terms of kinematical quantities on $K$,
the functions $F_L$, $F_t$ appear. But if the ratio $c_{o-}/\,c_{o+}$ is
considered, {\it the functions }$F_L${\it , }$F_t${\it \ disappear}, and the
final result

\begin{equation}
\rho \equiv \frac{c_{o-}}{c_{o+}}=\frac{1+\beta }{1-\beta }  \label{ro}
\end{equation}
($\beta \equiv \omega R/c$) is obtained. Now, if we consider the class of
rotating disks having the same peripheral velocity $\omega R$ and arbitrary
centrifugal acceleration $a=\omega ^2R$, the observable quantity $\rho $
given by expression (\ref{ro}) is constant for increasing radius ($%
R\rightarrow \infty $) and decreasing acceleration ($a\rightarrow 0$). Here
the paradox arises. In fact, uniform motion at any speed whatsoever may be
thought of as the limit of the motion on the rim of a disk of infinite
radius and infinitesimal acceleration; this means that, in the limit case of
null acceleration, the neighborhood of the light source $\Sigma $ on the
disk should be indistinguishable from an inertial frame. As a consequence,
such a neighborhood should be related to $K$ by a standard Lorentz
transformation, and the ratio $\rho $ between the speeds of the forward and
backward moving light rays should be exactly 1. Selleri concludes that SRT
gives rise to a discontinuity in the function $\rho (a)$ for $a\rightarrow 0$%
, and claims that such a discontinuity is inadmissible ''because our
empirical knowledge about inertial systems was actually obtained in frames
with small but non zero acceleration, e.g. because of Earth rotation''.

Since the calculations of Selleri are quite careful, the paradox cannot be
avoided until the assumption (a), which states that the round trip on the
turntable corresponds to a well defined circumference whose length is
univocally defined, is maintained. Of course, this paradox is lethal for the
self-consistency of the SRT; as a consequence, Selleri suggests that the SRT
should be abandoned and that the isotropy of the speed of light exists in
only one privileged reference frame, according to an idea he already
proposed elsewhere \cite{selleri1}.

Once again, let us stress that, according to proposition (b), eq. (\ref{ro})
does not depend on the particular choice of the functions $F_L$ and{\it \ }$%
F_t${\it ; }that's why{\it \ } eq. (\ref{ro}) coincides with the classical
result, corresponding to the Galilean velocity composition rule (which can
be obtained assuming $F_L=${\it \ }$F_t=1$).

{\it Remark: the anisotropy of the light velocity and the ''hypothesis of
locality''.}

{\it \ }Selleri stresses that eq. (\ref{ro}) ''does not only give the ratio
of the two global light velocities for a full trip around the platform in
the two opposite directions, but also the local ratio as well''.{\it \ }
This is consistent with the assumed symmetries of the disk, in particular
with the assumption of homogeneity of space along the rim; but conflicts
with the ''hypothesis of locality'' \cite{reichenbach},\cite{mashhoon}%
\footnote{%
''Hypothesis of locality'' is the expression used by Mashhoon to name one of
the most important axioms of Relativity Theory, which states the local
equivalence of an accelerated observer with a momentarily comoving inertial
observer (provided standard clocks and rods are used).}, according to which
the speed of light, as measured{\it \ locally} by means of standard rods and
clocks at rest in $K_{o\text{ }}$ (in an infinitesimal neighborhood of the
light source $\Sigma $) should be exactly the same as that observed in the
local inertial frame, the latter being $c$ in both directions.

{\it Remark: eq.(\ref{ro}) in some previous relativistic approaches.}

{\it \ }Eq. (\ref{ro}) has actually been obtained by many authors in
apparently relativistic contexts. Landau and Lifshitz \cite{landau}, \S\ 89,
and some other authors - more or less explicitly: see e.g. Arzeli\`{e}s \cite
{arzelies}, \S\ 115 - derived formulas equivalent to (\ref{ro}) at first
order in $\beta $. The underlying idea is that, since no transitive
synchronization procedure exists for the turntable in motion, the best time
to be introduced on the disk is not the (proper) time measured by real
clocks on it, but the time of the inertial frame $K$ (sometimes called
''universal time'' \cite{landau} or, more appropriately, ''central time'' 
\cite{arzelies}). From an operational point of view, this definition of time
means that any clock on the disk should not show its proper time, but the
time of the clock of $K$ over which it happens to be located at a given
instant \footnote{%
This is equivalent to a rescaling of the proper time at any point on the
disk by a factor of $\gamma ^{-1}$.}. This way the simultaneity criterium in 
$K_{o}$ is borrowed from $K$. As a consequence: (i) the spatial section of
the reference frame of the disk is the (Euclidean) 2-plane $x^{0}\equiv
ct=const,\,x^{3}\equiv \,z=const$; (ii) the coordinate transformations
between $K$ and $K_{o}$ take on the following Galilean-type form: 
\begin{equation}
\left\{ 
\begin{array}{c}
x_{o}^{0}=x^{0} \\ 
x_{o}^{1}=x^{1} \\ 
x_{o}^{2}=x^{2}+\omega t=x^{2}+\frac{\omega }{c}x^{0} \\ 
x_{o}^{3}=x^{3}
\end{array}
\right.  \label{trasf}
\end{equation}

where $x^1,\,x^2,x^3$ and $\,x_o^1,\,x_o^2,x_o^3$ are the cylindrical
coordinates $r,\,\theta ,z$ and $r_o,\,\theta _o,z_o$ in $K,\,K_o$
respectively.

Since the total round trip times for co-rotating and counter-rotating light
signals are not the same, {\it but the length of the two round trips in }$%
K_o ${\it \ is assumed to be the same (as being related to a univocally
defined geometric object)}, the velocity of light should be different for
the two patterns, and eq. (\ref{ro}) follows. Such a velocity seems to be
considered by Landau and Lifshitz \cite{landau} as a physical quantity,
because the physical time $t_o$ of $K_o$ differs from the ''universal time'' 
$t$ of $K$ only for terms of second order in $\beta $.

The same result (\ref{ro}) is obtained by Peres \cite{peres} in the full
SRT, under the same underlying assumptions. It is unclear whether the
velocity of light is just a coordinate velocity without a physical meaning,
or is actually physical: in this case it is tacitly assumed $F_L=${\it \ }$%
F_t=1$, and the relativistic approach is only apparent.

Anyway, the point we would stress is that, {\it in any case}, every
(classical or ''relativistic'') approach, based on the hypothesis that the
length of a round trip is related to a univocally defined geometric object,
unavoidably leads to eq. (\ref{ro}); but no authors, before Selleri,
realized its paradoxical consequences.

\section{Geometry of motion in 2+1 dimensions}

Let $K_o$ be a rigid circular platform, rotating with constant angular
velocity $\omega $ with respect to an inertial frame $K$, bearing a coaxial
cylindrical mirror with a light source $\Sigma $ just inside the mirror
surface. Two light rays are sent by the source in opposite directions along
the surface of the mirror. All the masses are assumed to be negligible.

What happens is easily described in 2+1 dimensions. For convenience reasons,
we shall use polar coordinates $x^0\equiv ct,$ $x^1\equiv r,\,x^2\equiv $ $%
\theta ,\,$ , where $r$ is the distance from the rotation axis and $\theta $
the rotation angle, as measured in $K$. In these coordinates, the metric
tensor takes the simple form 
\[
g_{\mu \nu }=\left( 
\begin{array}{ccc}
1 & 0 & 0 \\ 
0 & -1 & 0 \\ 
0 & 0 & -r^2
\end{array}
\right) 
\]

If $R$ is the value of $r$ for the source $\Sigma $ (and also the radius of
the mirror), the world line of $\Sigma $ (which is placed in a fixed point
on the platform) is a timelike helix, say $\gamma _\Sigma $, that wraps
around the cylinder representing the disk in $2+1$ dimensions. The
parametric equations of $\gamma _\Sigma $, in the coordinates $x^0\equiv ct,$
$x^1\equiv r,\,x^2\equiv $ $\theta ,\,$ of $K$, are 
\[
\left\{ 
\begin{array}{c}
x^0\equiv ct \\ 
x^1\equiv r=R \\ 
x^2\equiv \theta =\omega t
\end{array}
\right. 
\]

Eliminating the parameter $t$, they become 
\begin{equation}
\left\{ 
\begin{array}{c}
x^0=\frac c\omega \theta \\ 
x^1=R
\end{array}
\right.  \label{gammasigma}
\end{equation}

Notice that the length of the helix $\gamma _\Sigma $ is an observable
quantity, namely (apart a factor $c$) the proper time measured by a clock $%
C_\Sigma $ carried by the source $\Sigma $. Such length, expressed in units
of time, is 
\begin{eqnarray}
d\tau &=&\frac 1c\sqrt{g_{\mu \nu }dx^\mu dx^\nu }=\frac 1c\sqrt{%
c^2dt^2-R^2d\theta ^2}=  \label{elica} \\
&=&\frac 1c\sqrt{\frac{c^2}{\omega ^2}d\theta ^2-R^2d\theta ^2}=\frac{%
d\theta }\omega \sqrt{1-\beta ^2}  \nonumber
\end{eqnarray}

We see the time dilation at work.

Two light beams, emitted at the same time $t=0$ when the source $\Sigma $
(physically realized by means of two instruments, an electromagnetic source
and a beam splitter) is in $(0,R,0)$ (coordinates of $K$), travel along two
world lines which are null helixes, say $\gamma _{L_{\pm }}$ (the $+$ sign
holds for the co-rotating ray, the $-$ sign for the counter-rotating one).

The parametric equations of the helixes $\gamma _{L_{\pm }}$, in the
coordinates $x^0\equiv ct,$ $x^1\equiv r,\,x^2\equiv $ $\theta ,\,$ of $K$,
are 
\[
\left\{ 
\begin{array}{c}
x^0\equiv ct \\ 
x^1\equiv r=R \\ 
x^2\equiv \theta =\pm \frac cRt
\end{array}
\right. 
\]
from which, eliminating the parameter $t$: 
\begin{equation}
\left\{ 
\begin{array}{c}
x^0=\pm R\theta \\ 
x^1=R
\end{array}
\right.  \label{gammaluce}
\end{equation}

The two rays meet the source again at two different events which correspond
to the intersections between the null helixes $\gamma _{L_{\pm }}$ and the
timelike helix $\gamma _\Sigma $.

The (first) intersection between $\gamma _{L_{+}}$ and $\gamma _\Sigma $
(''absorption of the co-rotating photon by a detector placed in the same
place of the source $\Sigma $, after a complete round trip'') is found when 
\begin{equation}
\frac c\omega \theta =R\left( \theta +2\pi \right)  \label{uno}
\end{equation}

Analogously, the (first) intersection between $\gamma _{L_{-}}$ and $\gamma
_\Sigma $ (''absorption of the counter-rotating photon after a complete
round trip'') is found when 
\begin{equation}
\frac c\omega \theta =-R\left( \theta -2\pi \right)  \label{due}
\end{equation}
(of course the angle $\theta =\omega t$ is the rotation of $\Sigma $ in the
time interval $t$, everything being measured in $K$).

Eq.s (\ref{uno}), (\ref{due}) show that the two photons, emitted from the
source $\Sigma $ at the time $t=0$ and travelling in opposite directions,
are absorbed - after a complete round trip around the rim of the platform -
by the detector, placed in the same place of the source, when the angular
coordinates of $\Sigma $ - as measured in $K$ - are, respectively, 
\begin{eqnarray}
\theta _{\Sigma _{+}} &=&\frac{2\pi \beta }{1-\beta }  \label{angoli} \\
\theta _{\Sigma _{-}} &=&\frac{2\pi \beta }{1+\beta }  \nonumber
\end{eqnarray}
Introducing these results into eq.(\ref{elica}), we see that the two
''absorption events'' happen at the following proper times of $\Sigma $
(times measured by the standard clock $C_\Sigma $, at rest on the platform
in $\Sigma $): 
\begin{eqnarray}
\tau _{+} &=&\frac{\theta _{\Sigma _{+}}}\omega \sqrt{1-\beta ^2}=\frac{2\pi
\beta }\omega \sqrt{\frac{1+\beta }{1-\beta }}  \label{tempo+-} \\
\tau _{-} &=&\frac{\theta _{\Sigma _{-}}}\omega \sqrt{1-\beta ^2}=\frac{2\pi
\beta }\omega \sqrt{\frac{1-\beta }{1+\beta }}  \nonumber
\end{eqnarray}
and are separated by the proper time interval: 
\begin{equation}
\delta \tau \equiv \tau _{+}-\tau _{-}=\frac{\theta _{\Sigma _{+}}-\theta
_{\Sigma _{-}}}\omega \sqrt{1-\beta ^2}=\frac{4\pi \beta ^2}{\omega \sqrt{%
1-\beta ^2}}  \label{properlag}
\end{equation}

Taking into account the time dilation in the inertial frame $K$, the
corresponding universal time interval $\delta t$ is: 
\begin{equation}
\delta t=\gamma \delta \tau =\frac{4\pi \beta ^2}{\omega \left( 1-\beta
^2\right) }  \label{lag}
\end{equation}

{\it Remark: rotation angles for the light rays.}

{\it \ }Let $\theta _{L_{+}}$, $\theta _{L_{-}}$ be the rotation angles (as
measured in $K$) of the co-rotating and counter-rotating light beams when
they are absorbed by the detector after a complete round trip. Then from
eqs. (\ref{angoli}): 
\begin{eqnarray}
\theta _{L_{+}} &=&\theta _{\Sigma _{+}}+2\pi =\frac{2\pi }{1-\beta }
\label{angoli1} \\
\theta _{L_{-}} &=&\theta _{\Sigma _{-}}-2\pi =-\frac{2\pi }{1+\beta } 
\nonumber
\end{eqnarray}

{\it Remark: the Sagnac effect. }

Eqs. (\ref{properlag}), (\ref{lag}) exactly coincide with the formulas which
are at the basis of the Sagnac effect, i.e. the eqs. (23), (22) of Post \cite
{post} (see also Stedman \cite{sagnac}, Dieks \cite{dieks} and, as it will
be more apparent in the next section, Anandan \cite{anandan}).

As known, the Sagnac effect is a shift of the interference fringes appearing
in a suitable interferometer, and is due to the time difference (given
either by eq. (\ref{properlag}) or by eq. (\ref{lag}), according to the
particular choice of the clock) between the arrivals on the detector of the
co-rotating and the counter-rotating light beam.

The classical explanation (see for instance Sagnac \cite{sagnac1913}), but
also many ''relativistic'' explanations (e.g. Peres \cite{peres}), ascribe
such a time difference to the anisotropy of light propagation along the rim
of the platform, due to rotation.

On the other hand, the true relativistic explanation, proposed by Anandan 
\cite{anandan} in 1981 and here recovered, with many interesting
implications, ascribes such a time difference to the nonuniformity of time
on the rotating platform, and in particular to the ''time lag'' arising in
synchronizing clocks along the rim (see next section).

\begin{figure}[ph]
\caption{General view of the geometry of the rotating disk in 2+1 dimension.
A few timelike helixes are drawn, belonging to the congruence which defines
the reference frame $K_0$ of the disk. The dashed line $\gamma _{L+}$
represents the co-rotating light beam.}
\end{figure}

\section{Lengths along the rim of the disk}

In order to compare the speed of the light beams as seen on board the
platform, the lengths of the different travels are needed. To this end,
consider a given event taking place on the rim of the platform, e.g. the
event ''emission of two photons in opposite directions from the source $%
\Sigma $''. This event will be denoted by the symbol $\Sigma \left( 0\right) 
$ partly recalling the one already used for the source/detector (of course
the two meanings of $\Sigma $ should be clear from the context, in order to
avoid any confusion).

The locus of events ''simultaneous to $\Sigma \left( 0\right) $'' is defined
without ambiguities in the class of the time-orthogonal and geodesic
reference frames (see \cite{cattaneo}), which contains the class of the
(extended or local) inertial{\it \ } reference frames.

In particular, the set of simultaneous events is univocally defined: (i) in
the {\it extended} inertial frame $K$ ; (ii) in an infinitesimal region of
the platform, containing $\Sigma $, which differs as little as we want from
the {\it local} inertial frame $K_{o}\left( \Sigma \right) $. When we start
from the event $\Sigma \left( 0\right) $ and move along the rim of the
platform, the simultaneity procedure, as defined in $K_{o}\left( \Sigma
\right) $, is transported, step by step, along the rim. As a result, the set
of events taking place on the rim and simultaneous to $\Sigma $ in $K_{o}$
(i.e. satisfying the condition $x_{o}^{0}=0$) is mapped, in the
(3-dimensional Euclidean) plot of the (2+1) Minkowskian space-time, into a
spacelike helical curve $\gamma _{S}$, starting from $\Sigma \left( 0\right) 
$ and everywhere orthogonal to the timelike helix $\gamma _{\Sigma }$
(''history of the point $\Sigma $ at rest on the disk''), whose tangent
vector forms a constant angle $\alpha $ with respect to the $x^{0}$ axis of $%
K$. Orthogonality is shown in the plot (see figure 2) by the fact that the
angle between the tangent vector to $\gamma _{S}$ and a normal section of
the cylinder on which the helixes wrap is again everywhere the same $\alpha $%
\footnote{%
As it is well known, a spacetime diagram is a topological map from a
Minkowskian space to an Euclidean space, which changes the metrical
relations (angles and lengths) in a well defined way; in particular, the
M-orthogonality between two directions is depicted in the diagram as an
E-symmetry with respect to the light cone.} : 
\[
\alpha =\arctan \left( \beta \right) \,\,\,\, 
\]

Notice that, when the platform does not rotate ($\beta =0$), the world line
of any point on its border is a straight vertical line and the locus of the
events simultaneous to $\Sigma $ is a {\it closed} curve, namely a
circumference (whose length coincides with the usual length of the contour
of the disk); but when the platform is set in motion ($\beta \neq 0$), the
world lines of the points on its border become timelike helixes\footnote{%
If the totality of the points of the platform is considered, then the
totality of their world lines is a congruence of timelike helixes, which
should be assumed as the only unambigous definition of the reference frame $%
K_o$ (see \cite{cattaneo}).} and the sets of simultaneous events change
completely their topology and become {\it open} curves, namely spacelike
helixes orthogonal to the former ones.

In particular, the helix $\gamma _S$ of the events simultaneous to $\Sigma
\left( 0\right) $ in $K_o$ is easily found by imposing that the tangent unit
vector ${\bf \upsilon }_{\gamma _S}=\left( \upsilon _{\gamma _S}^0,\upsilon
_{\gamma _S}^\theta \right) $ to $\gamma _S$ be normal to the tangent vector
to $\gamma _\Sigma $. One obtains 
\begin{eqnarray}
\upsilon _{\gamma _S}^0 &\equiv &c\frac{dt}{ds_{\gamma _S}}=\frac \beta {%
\sqrt{1-\beta ^2}}  \label{tg-gammas} \\
\upsilon _{\gamma _S}^\theta &\equiv &\frac{d\theta }{ds_{\gamma _S}}=\frac
1{R\sqrt{1-\beta ^2}}  \nonumber
\end{eqnarray}
from which, dividing the second formula by the first one, then integrating,
the following equations for $\gamma _S$ follow: 
\begin{equation}
\left\{ 
\begin{array}{c}
r=R \\ 
\theta =\frac{c^2}{\omega R^2}t
\end{array}
\right.  \label{gammas}
\end{equation}

Considering (\ref{gammas}) the infinitesimal length along the helix $\gamma
_S$ is 
\begin{eqnarray}
ds_{\gamma _S} &=&\sqrt{g_{00}\left( dx^0\right) ^2+g_{\theta \theta }\left(
dx^\theta \right) ^2}=\sqrt{c^2dt^2-R^2d\theta ^2}  \label{dsgamma} \\
&=&\sqrt{\omega ^2\frac{R^4}{c^2}d\theta ^2-R^2d\theta ^2}=iRd\theta \sqrt{%
1-\beta ^2}  \nonumber
\end{eqnarray}

If valued in $K_{o}$, the line element $ds_{\gamma _{S}}$ should be
interpreted, leaving the imaginary unit $i$ aside\footnote{%
The appearance of the imaginary unit is due to the conventions regarding the
signature of the four dimensional line element and simply means that the
interval is spacelike.}, as the ''proper length'' $\left( dl\right) _{o}$ of
an infinitesimal part of the rim in its locally inertial frame: 
\begin{equation}
\left( dl\right) _{o}=Rd\theta \sqrt{1-\beta ^{2}}  \label{dlo}
\end{equation}
Of course, eq. (\ref{dlo}) has a univocally well defined interpretation only
for an infinitesimal part of the rim; if we integrate along a finite portion
of it, the interpretation of such integral as the ''proper length of the
considered part of the rim'' is a questionable extrapolation. In fact, a
rotating disk does not admit a well defined ''proper frame''; rather, it
should be regarded as a class of an infinite number of local proper frames,
considered in different points at different times, and glued together
according to some convention. It is well known (\cite{cantoni}, \cite
{anandan}, \cite{arzelies}, \cite{landau}, etc.) that no convention exists
such that a self-consistent synchronization of standard clocks at rest on
the disk can be realized (see later on). In particular, the (first)
intersection $\Sigma \left( \tau _{0}\right) $ (see fig.2) of the helix $%
\gamma _{\Sigma }$ - given by eq.(\ref{gammasigma}) - with the helix $\gamma
_{S}$ of the events ''simultaneous to $\Sigma \left( 0\right) $ in $K_{o}$''
- given by eq.(\ref{gammas}) - takes place when the angular coordinate (in $%
K $) of $\Sigma $ takes on the value $\theta _{\Sigma }$ satisfying the
equation $\frac{\theta _{\Sigma }}{\omega }=\left( 2\pi +\theta _{\Sigma
}\right) \frac{\omega R^{2}}{c^{2}}$, namely 
\[
\theta _{\Sigma }=2\pi \beta ^{2}\left( 1-\beta ^{2}\right) ^{-1} 
\]
Along the ''simultaneity helix'' $\gamma _{S}$, the intersection $\Sigma
\left( \tau _{0}\right) $ takes place after the rotation angle $\theta
_{0}=2\pi +\theta _{\Sigma }=2\pi \left( 1-\beta ^{2}\right) ^{-1}$. But the
event $\Sigma \left( \tau _{0}\right) $, though ''simultaneous to $\Sigma
\left( 0\right) $ in $K_{o}$'' according to the previous definition, {\it \
belongs to the future of }$\ \Sigma \left( 0\right) $! This is a well known
example which displays the impossibility of a self-consistent definition of
simultaneity {\it at large} on the disk.

Eq. (\ref{dlo}) deserves a further comment. The quantity $Rd\theta $
appearing in it cannot be interpreted as the length of an infinitesimal arc
of the rotating circumference as viewed in the inertial frame $K$, because
to come back to the source on the turntable the rotation angle must increase
by $\theta _{0}=2\pi \left( 1-\beta ^{2}\right) ^{-1}>2\pi $. If we want a
round trip on the platform to correspond to a $2\pi $ rotation, we need a
new angular variable $\theta ^{\prime }$ such that $d\theta ^{\prime }$ =$%
d\theta \left( 1-\beta ^{2}\right) .$ In terms of this new variable, eq. (%
\ref{dlo}) takes the form

\begin{equation}
\left( dl\right) _{o}=Rd\theta ^{\prime }\left( 1-\beta ^{2}\right) ^{-\frac{%
1}{2}}  \label{dlzero}
\end{equation}
Stated in other words, $Rd\theta $ and $Rd\theta ^{\prime }$ are the
projections of $\left( dl\right) _{o}$ onto a plane $t=$const of the
inertial frame $K$, along $x^{o}$ and $\gamma _{\Sigma }$ respectively. As a
consequence, only the latter expression should be interpreted as the length
of an infinitesimal arc of the rotating circumference as viewed in $K$, and
we recognize in the previous expression the Lorentz contraction.

In spite of the impossibility of a self-consistent definition of
simultaneity {\it at large} on the disk, it is interesting to stress that
the length of the rim in $K_o$, as defined by the formula 
\begin{equation}
l_o\equiv \int\limits_{\Sigma \left( 0\right) }^{\Sigma \left( \tau
_0\right) }\left( dl\right) _o=\int\limits_0^{\theta _0}Rd\theta \sqrt{%
1-\beta ^2}=2\pi R\left( 1-\beta ^2\right) ^{-\frac 12}  \label{lo}
\end{equation}
exactly coincides with the expected relation between the length $l_o$ of the
''circumference'' 
\begin{equation}
\gamma _o\equiv \left\{ P(\theta )\in \gamma _S:\,\theta \in \left[ 0,\theta
_0\right] \,\right\}  \label{gamma00}
\end{equation}
relative to $K_o$, and the length $l=2\pi R$ of the circumference relative
to $K$. However, the interpretation of this result differs from the
traditional one (whose origin can be found in Einstein \cite{einstein}): the
main issue, first remarked by Cantoni \cite{cantoni} and later on by Anandan 
\cite{anandan}, is that the ''circumference'' $\gamma _o$ is not a
circumference at all, but an {\it open} spacelike curve, whose end-point $%
\Sigma \left( \tau _0\right) $ belongs to the future of the starting-point%
{\it \ }$\Sigma \left( 0\right) $. The time distance between $\Sigma \left(
0\right) $ and $\Sigma \left( \tau _0\right) \,$along $\gamma _\Sigma $, as
measured by the clock $C_\Sigma $ at rest near $\Sigma $, is the ''time
lag'' 
\begin{equation}
\tau _0=\frac{\theta _0}\omega \sqrt{1-\beta ^2}=\frac{2\pi \beta ^2}{\omega 
\sqrt{1-\beta ^2}}  \label{timelag}
\end{equation}
(see eq. (\ref{elica})) which arises in synchronizing clocks around the rim,
because of the rotation.

Notice that the ''time lag'' (\ref{timelag}) is exactly half of the proper
time interval (\ref{properlag}) between the arrivals on the detector of the
co-rotating and the counter-rotating light beams: this means that the Sagnac
effect can be explained as a consequence of the ''time lag'' due to rotation
(see \cite{anandan}).

But the main conclusion of this paper is the following: the two light beams
come back to the detector (placed near the source $\Sigma $) {\it for
different values of the rotation angle around the cylinder}, as given by eq.
(\ref{angoli1}). Actually, the two end-points $\Sigma \left( \tau
_{+}\right) \equiv \Sigma (\theta _{L_{+}}=\theta _{\Sigma _{+}}+2\pi
),\,\Sigma \left( \tau _{-}\right) \equiv \Sigma (\theta _{L_{-}}=\theta
_{\Sigma _{-}}-2\pi )$ along $\gamma _\Sigma $ are different; more
explicitly, there are two different travelled distances for the two light
rays leaving $\Sigma \left( 0\right) $ in opposite directions, according to 
\begin{eqnarray}
l_{+} &\equiv &\int\limits_{\Sigma \left( 0\right) }^{\Sigma \left( \tau
_{+}\right) }\left( dl\right) _o=\int\limits_0^{\theta _{L_{+}}}Rd\theta 
\sqrt{1-\beta ^2}=2\pi R\sqrt{\frac{1+\beta }{1-\beta }}  \label{lungh-luce}
\\
l_{-} &\equiv &\int\limits_{\Sigma \left( 0\right) }^{\Sigma \left( \tau
_{-}\right) }\left( dl\right) _o=\int\limits_0^{\left| \theta
_{L_{-}}\right| }Rd\theta \sqrt{1-\beta ^2}=2\pi R\sqrt{\frac{1-\beta }{%
1+\beta }}  \nonumber
\end{eqnarray}

Now the effective speeds of light on the platform, in the two opposite
directions, are given by the ratios of the two travelled lengths $%
l_{+},\,l_{-}$ to the proper travel times as measured by the clock $C_\Sigma 
$ at rest in $\Sigma $, which are given by eqs. (\ref{tempo+-}). The result
is 
\begin{eqnarray}
c_{+} &\equiv &\frac{l_{+}}{\tau _{+}}=\frac{R\omega }\beta =c
\label{finale} \\
c_{-} &\equiv &\frac{l_{-}}{\tau _{-}}=\frac{R\omega }\beta =c  \nonumber
\end{eqnarray}

Both these velocities, that correspond to what Selleri calls $\widetilde{c}%
\left( 0\right) $ and $\widetilde{c}\left( \pi \right) $, are exactly equal
to $c$; therefore their ratio is simply $1$ just as in inertial reference
frames. As a consequence: (i) no discontinuity is found in passing from
accelerated to uniform motion; (ii) no violation of the ''hypothesis of
locality'' occurs; (iii) the explanation of the Sagnac effect does not
require any anisotropy in the propagation of light along the rim, as often
claimed \cite{peres}, but is {\it totally }due to the time-lag (\ref{timelag}%
) arising in synchronizing clocks around the rim, because of the rotation.

\begin{figure}[tbp]
\caption{The figure shows the $2+1$ diagram of the rim of the rotating disk
developed on a plane: vertically the time $t$ of reference $K$ $is$ shown;
horizontally one finds the rotation angles as seen in $K$. Primed and
unprimed low case letters on opposite sides of the figure mark one and the
same point on the cylinder. The symbols are the same as those used in the
text. Simple geometric properties reproduce the results obtained in the
text. Null lines are plotted at $45^{o},$ thus it is immediately seen that $%
\tau _{0}$, distance between $\Sigma \left( \tau _{0}\right) $ and $\Sigma
\left( 0\right) ,$ is half $\tau _{+}-\tau _{-}$ (distance between $\Sigma
\left( \tau _{+}\right) $ and $\Sigma \left( \tau _{-}\right) $. The length
of $\gamma _{o}$ is the length of the line $\Sigma \left( 0\right)
bb^{\prime }\Sigma \left( \tau _{0}\right) $, $l_{+}$ is $\Sigma \left(
0\right) bb^{\prime }n$; $l_{-}$ is $\Sigma \left( 0\right) a^{\prime }am$,
which is the same as $\Sigma \left( 0\right) bb^{\prime }m^{\prime }$:
indeed $m^{\prime }$ is the first point on the world line of the
counter-rotating light ray ''simultaneous'' to $\Sigma \left( 0\right) $.}
\end{figure}

\section{Comparison with a ''natural'' splitting between space and time}

In 4 dimensions everything is clear and clean, but we perceive the world
locally in 3+1 dimensions; so the results found in the previous sections
could be considered as being rather abstract requiring the knowledge and
application of four-dimensional geometry and setting aside what appears to
be a most ''natural'' splitting between space and time. The question can
legitimately be posed whether the same general result could be found on the
base of actual measurements performed on the disk.

In the present section, we compare the four-dimensional treatment, outlined
in sect. 3 and sect. 4, with the 3+1 global splitting suggested, in an
apparently ''natural'' way, by a pure experimenter living on the platform,
on the basis of actual measurements outside of any four-dimensional
interpretation.

Our observer possesses a great many unitary rods and a couple of identical
clocks. The clocks are synchronized in one and the same place, that will be
assumed to be the origin of the observer's reference frame.

Now the observer sets out for an extremely slow (negligible relative
velocity) round trip on the platform in a sense which for us is corotating,
carrying with him the rods and one of the clocks. At each step he lays down
one rod, tail to head of the previous one. When he comes back to the origin
the number of rods he used tells him what reasonably he considers as being
the length $l_0$ of his trip. There is however a curious phenomenon our
experimenter can notice: when coming back to the origin, the clock carried
with him turns out to be desynchronized with respect to the identical one he
left at the origin, and precisely turns out to be {\it late} by a given
amount $\tau _0$, whose numerical value is given by eq. (\ref{timelag}).

If our observer repeats his slow trip in the reverse sense, he will find:
(i) the same length $l_0$ of his new trip, since the number of rods is one
and the same both turning clockwise and counterclockwise; (ii) the same
desynchronization of his clock, which this time turns out to be {\it ahead}
by the same amount $\tau _0$ with respect to the identical one at the
origin. If he compares the results of the two trips, he finds a net
desynchronization $2\tau _0$\ between the two travelling clocks, after their
clockwise and counterclockwise slow round trips.

To solve this puzzling result, our experimenter decides to send light beams
in both directions along the same path he followed and measures the time
they take to reach the origin again, finding two different results $\tau _{+}
$ and $\tau _{-}$. The most interesting thing is that {\it the difference }$%
\tau _{+}-${\it \ }$\tau _{-}${\it \ exactly equals the net
desynchronization }$2\tau _{0}${\it \ between the two (slowly) travelling
clocks!} Excited by this astounding coincidence, our friend decides to send
pairs of material objects (particles or matter waves) in opposite
directions, with the same relative velocity. He finds of course different
times of the total round trips, but the same difference between them for any
pair of travelling objects: $2\tau _{0}$ again and again!\footnote{%
This result, although known in the literature, is not demonstrated in the
present paper, but will be confirmed (and perhaps more clearly established)
in a brief letter in preparation.} So our experimenter, although completely
unaware of any theoretical approach to the problem, in particular of any
four-dimensional geometrical model of space-time, and only confident in his
own measurements realized by means of real rods and real clocks, is forced
to conclude that: (i) the platform on which he lives is rotating, and the
desynchronization $2\tau _{0}$ of a pair of clocks, after their slow round
trips in opposite directions, is a measure of the speed of this rotation;
(ii) the time intervals along the closed path are not uniquely defined and,
to obtain reliable measures of them, the readings of clocks must be
corrected by a quantity $\pm \tau _{0}$ to account for the desynchronization
effect (by the way, this is precisely what is done when considering data
from the GPS\footnote{%
Global Positioning System} satellites \cite{allan}); (iii) as a consequence
of this correction, the speed of light is actually the same both forward and
backward.

It is interesting to note that our pure experimenter discovers the
interdependence between space and time, although unaware of the metric
structure of Minkowski space-time - or, what is the same - of SRT.

\bigskip

To sum up, we see that different approaches are possible about space and
time, however once one has been chosen about space (time), the properties of
time (space) are entirely determined by the theory and produce in any case
the same result. Different pictures are available but the statements about
physical quantities and their relationships are finally the same.

(1) According with our pure experimenter, the circumference of the rotating
disk can be considered as a geometrically well defined entity, with a well
defined length in the reference frame of the disk. But in this case, when
considering durations involving displacements along the rim (i.e. through
points where time flows differently), {\it a correction is unavoidable} {\it %
in order to take into account the phenomenon of desynchronization}. This is
the price to be payed if consistent results are desired.

(2) According to our four-dimensional treatment, the synchronization along
the rim is defined by extending formally the local Einstein criterium of
simultaneity. Then the length of the circumference of the rotating disk, as
measured by infinitesimal rigid rods at rest along the rim, turns out to be
an open curve in space-time, and its length turns out to be
traveller-dependent.

One can choose description (1), which is closer to real measurements on the
platform, or description (2), which is clearer and, at least in our opinion,
quite enlightening; but the conclusions about physical quantities are
exactly the same. The choice of the description is a matter of taste; the
statements about physical quantities and their relationships are a matter of
fact and self-consistency.

\section{Conclusion}

The starting point of this paper was a careful check of eq. (\ref{ro}). We
easily found that that formula, widespread both in non-relativistic and in
relativistic literature, and apparently supported by the experimental
evidence of the Sagnac effect, actually leads to the paradoxical
consequences pointed out by Selleri \cite{selleri}. In particular, eq. (\ref
{ro}) turns out to be incompatible both with the principle of invariance of
the speed of light in the class of the inertial frames and with the
''hypothesis of locality''. As a consequence, eq. (\ref{ro}) would rule out
all the relativistic physics of this century and imply the existence of a
privileged (''absolute'') frame, which could be interpreted as the frame of
the stationary ether. We are not going to comment further on some unexpected
and radical implications, first of all the recovery of absolute
simultaneity; see Sagnac \cite{sagnac1913} for a Galilean-like
interpretation, and Selleri \cite{selleri}, \cite{selleri1} for a
Lorentzian-like interpretation.

Actually we have shown that eq.(\ref{ro}) comes from the (explicit or
implicit) assumption that the length of round trip journeys along the rim of
a turntable is that of a closed curve (namely a circumference) and is unique
for all travellers independently from the kind of synchronization procedure
adopted. This assumption, evident in eq.(\ref{c}), where the same length $%
L_{o}$ is used for both the counter-rotating and the co-rotating light
signals (as seen in $K_{o}$), comes from a purely three-dimensional approach
(global $3+1$ splitting). However if we look at relativity from a
4-dimensional (in this case $2+1$ dimensional is enough) point of view, one
sees that ''round trips'' correspond in general to open curves (arcs of
helixes) and their proper lengths differ from one traveller to another, in
particular for co-rotating and counter-rotating light beams (the starting
point $\Sigma (0)$ is univocally defined, but the end point $\Sigma (\tau )$
depends on $\tau $, as can be seen on figure 2). When the speed of the
traveller tends to zero (in any direction) the length of the journey tends
to the unique value $l_{0}$ given by eq.(\ref{lo}), which has usually been
considered as the length of the ''circumference'' $\gamma _{0}$ defined in (%
\ref{gamma00}).

The root of the misunderstandings lies in the ambiguity of a self-consistent
definition of simultaneity (in $K_{o}$) of events taking place along the rim
of the disk, in particular in the drastic change of the topology of the line
of simultaneous events, due to rotation, which is evident in a
four-dimensional context only. On this very fact (namely in the time lag of
eq.(\ref{timelag})) stands the correct relativistic explanation of the
Sagnac effect, which does not need any anisotropy of the speed of light
along the rim, contrary to a widely supported idea.

In order to formally recover, in that context, the isotropy of the speed of
light, Peres is forced to introduce an ad hoc time that has nothing to do
with the time of real clocks on the turntable \cite{peres}. Our result, in a
full Minkowskian context, is that the speed of light is actually the same
both ''eastward and westward'' (to use Peres' words), as measured using real
clocks at rest on the platform.

In conclusion, we have shown that the SRT has no flaws when applied to
describe the behaviour of light as seen from a turntable carrying mirrors,
provided we avoid the use of (geometrical or kinematical) quantities {\it %
ambiguously defined}, and stick consistently to its axioms and rules.

\end{document}